\documentstyle[11pt,epsf]{article}
\def\be{\begin{equation}}
\def\ee{\end{equation}}
\def\bea{\begin{eqnarray}}
\def\eea{\end{eqnarray}}

\newtheorem{theorem}{Theorem}
\newtheorem{prop}{Proposition}
\newtheorem{corr}{Corollary}
\newtheorem{lemma}{Lemma}

\def\eqq{\stackrel{\sigma}{=}}

\def\sup{\sigma}
\def\suup{\sigma}
\def\mmm{{\cal V}}
\def\mm2{{\cal V}}
\def\d{d}

\newfont{\BlackBoardBold}{msbm10}
\def\RR{{\rm I\kern-.17em R}}

\def\be{\begin{equation}}
\def\ee{\end{equation}}
\def\bea{\begin{eqnarray}}
\def\eea{\end{eqnarray}}

\def\p{\partial}

\def\a{\alpha}

\def\b{\beta}

\newcommand\bm[1]{\mbox{\boldmath $#1$}}
\begin{document}
\title{Matching of spatially homogeneous non-stationary space--times
to vacuum in cylindrical symmetry}
\author{Paul Tod$^{1}$ \& Filipe C. Mena$^{1,2}$\\
{\em $^1$
Mathematical Institute,
University of Oxford,
St. Giles 24-29, Oxford OX1 3LB, U.K.}
\\
{\em $^2$
Departamento de Matem\'atica,
Universidade do Minho, Gualtar,
4710 Braga, Portugal}\\}
\date{10th May, 2004}
\maketitle
\begin{abstract}
We study the matching of LRS spatially homogeneous collapsing dust
space-times with non-stationary vacuum exteriors in cylindrical
symmetry. Given an interior with diagonal metric we prove
existence and uniqueness results for the exterior. The matched
solutions contain trapped surfaces, singularities and Cauchy
horizons. The solutions cannot be asymptotically flat and we
present evidence that they are singular on the Cauchy horizons.
\end{abstract}
\section{Introduction}
In cylindrical symmetry, unlike spherical symmetry, there exist
exact solutions of the Einstein field equations (EFEs) which
contain gravitational radiation. This raises the possibility of
constructing collapsing matter cylinders matched to vacuum but
radiating exteriors. In particular one can attempt to find
analogues in cylindrical symmetry of the Oppenheimer-Snyder model.
The Oppenheimer-Snyder model results from the matching of a ball in the
Friedmann-Lemaitre-Robertson-Walker (FLRW) dust space-time to a
Schwarzschild exterior across a timelike surface preserving the
spherical symmetry. Thus there is no radiation in the exterior.
This model cannot be generalised (in a physically reasonable way)
to cylindrical symmetry with spatially homogeneous interiors and
static exteriors. In fact, Mena, Tavakol and Vera \cite{MTV} have
shown that the matching of such space-times leads to very
restrictive interior solutions with no evolution along one
spacelike direction. As a next step one  therefore considers a
non-static exterior, but for simplicity retains a spatially
homogeneous interior.

If such a model is to represent a cylinder of dust collapsing in a
vacuum, then it should be asymptotically flat. Berger, Chrusciel
and Moncrief \cite{BCM} have shown that for cylindrically
symmetric vacuum spacetimes the requirement of asymptotic flatness
implies that the $2$-surfaces of transitivity are never trapped
(specifically they show that one can choose a comoving radius of
the form $R(T,\rho)=\rho$, which forbids trapping).

In this paper we investigate a family of models which result from
the matching of a spatially homogeneous collapsing dust interior
with a non-stationary vacuum exterior. These interiors always
contain cylinders of symmetry (which we shall call
just `cylinders' for brevity) which are trapped near to the
singularity. The marginally-trapped cylinders trace out a
$3$-surface in the interior which eventually arrives at the
boundary of the matter. Matching conditions at the boundary mean
that the intersection $2$-surface at the boundary is
marginally-trapped as seen from the exterior and the results of
Berger et al. \cite{BCM} then prevent the exterior from being
asymptotically flat. We shall argue that the failure of asymptotic
flatness is in fact due to singularities in the vacuum exterior.

The technique will be to use the interior, which can be assumed to
be known, to give data on the boundary of the matter, which will
be a co-moving cylinder from the point of view of the interior but
just a time-like surface made up of cylinders of transitivity seen
from the outside. The EFEs in the exterior are a system of
hyperbolic equations in the $2$-dimensional quotient space
${\cal{Q}}$ of the space-time by the symmetries (this is the
$(T,\rho)$-space) and data can be given on any non-characteristic
curve in ${\cal{Q}}$. The boundary of the matter defines such a
curve, even though it is a time-like surface in the full
$4$-metric. Therefore we can deduce existence and uniqueness of
solutions in the domain of dependence in ${\cal{Q}}$ of the curve
on which data is given. If the interior collapses to a
singularity, as it does in these examples, then the data-curve has
an end and there is a Cauchy horizon in the exterior. One expects
the exterior to be singular on this Cauchy horizon, and we present
some evidence for this belief. We shall interpret the singularity
at the Cauchy horizon in terms of incoming gravitational
radiation.

The plan of the paper is as follows. In Section 2 we review the
matching procedure in general, and in Section 3 we introduce the
interior and exterior metrics which we wish to match. The matching
conditions for these metrics are written out in Section 4. In
Section 5, we give some properties of the matched space-times when
the metrics are diagonal in the canonical coordinates, obtaining
existence from standard theorems for linear PDEs. We briefly
discuss a non-diagonal example in Section 6 and finally in Section
7 we discuss some properties of the solutions, arguing that the
Cauchy horizon is singular and giving a physical reason for this.
Some equations are collected into an Appendix.

We use units such that $8\pi G=c=1$. Greek indices take values 1
to 3 and latin indices 0 to 3.
\section{Matching procedure in brief}
In this section, we recall how two space-times can be matched
across a hypersurface by showing which junction conditions the
respective metrics should satisfy. The first set of junction
conditions ensures the continuity of the metrics across the
matching hypersurface. The second set of junction conditions
ensures the continuity of the second fundamental forms and, given the first set,  is
equivalent to requiring a non-singular Riemann tensor distribution, which
 prevents infinite discontinuities of matter and curvature
across the matching hypersurface. In what follows we assume that
the matching surface is non-null. For cases where the matching
surfaces can change character see \cite{Mars-Seno}.

{}Let $(\mmm^+,g^+)$ and $(\mmm^-,g^-)$ be two $C^3$ space-times
with oriented boundaries $\sup^+$ and $\sup^-$, respectively, such that $\sup^+$ and $\sup^-$ are
diffeomorphic. The matched space-time
$(\mmm,g)$ is the disjoint union of $\mmm^\pm$
with the points in $\sup^\pm$ identified
 such that
the junction conditions are satisfied
(Israel \cite{Israel}, Clarke \& Dray \cite{Clarke-Dray} and
Mars \& Senovilla \cite{Mars-Seno}).
Since $\sup^\pm$ are diffeomorphic, one can then view those boundaries as
diffeomorphic to a
3-dimensional oriented manifold $\sup$ which can be
embedded in $\mm2^+$ and $\mm2^-$.
Let $\{\xi^\a\}$ and $\{x^{\pm i}\}$
be coordinate systems on $\sup$ and $\mm2^\pm$ respectively. The two embeddings
are given by the following $C^3$ maps
\bea
\label{embed}
\Phi^\pm:\; \sup &\longrightarrow& \mm2^\pm\\
\label{eq:embeddings}
             \xi^\a &\mapsto& {x^i}^\pm={\Phi^i}^\pm(\xi^\a),
\nonumber
\eea such that $\sup^\pm \equiv \Phi^\pm(\sup)\subset
\mm2^\pm$. The diffeomorphism from $\sup^+$ to $\sup^-$ is
$\Phi^-\circ {\Phi^+}^{-1}$.

Given the basis $\{\partial/\partial \xi^\a|_p\}$ of the tangent
plane $T_p \sup$ at some $p\in \sup$, the push-forwards
$d\Phi^\pm|_{p}$ map $\{\partial/\partial \xi^\a|_p\}$ into three
linearly independent vectors at $\Phi^\pm (p)$ represented by
$\vec e^{\;\pm}_\a|_{\Phi^\pm(p)}$:
\[ \label{push} \d
\Phi^\pm\left(\left.\frac{\partial}{\partial
\xi^\a}\right|_{\suup}\right) =\frac{\partial \Phi^{\pm
i}}{\partial \xi^\a} \left.\frac{\partial}{\partial x^{\pm
i}}\right|_{\suup^\pm} \equiv \vec
e^{\;\pm}_\a|_{{}_{\suup^\pm}}=e^{\pm i}_\a
\left.\frac{\partial}{\partial x^{\pm i}}\right|_{\suup^\pm}.
\nonumber
\]
On the other hand, using the pull-backs $\Phi^{\pm*}$ of the maps
$\Phi^\pm$, the metrics $g^\pm$ can be mapped to $\sup$ given two
symmetric 2-covariant tensors $\bar{g}^+$ and $\bar{g}^-$ whose
components in the basis $\{\d \xi^\a\}$ are \[ \label{hostia}
\bar{g}^\pm_{\a\b}\equiv e^{\pm i}_\a e^{\pm j}_\b
g_{ij}|_{{}_{\suup^\pm}}= (\vec e^{\;\pm}_\a\cdot \vec
e^{\;\pm}_\b)|_{{}_{\suup^\pm}}. \] The {\em first matching
conditions} are given by the equality of the first fundamental
forms (Israel \cite{Israel}) \be \label{aija5}
\bar{g}^+_{\a\b}=\bar{g}^-_{\a\b}. \ee We note that the existence
of a continuous metric allows for the treatment of Einstein's
field equations in the distributional sense (see e.g. Mars \&
Senovilla \cite{Mars-Seno}). The bases $\{\vec e^{\;+}_\a|_p\}$
and $\{\vec e^{\;-}_\a|_p\}$ can be identified, \[ \d
\Phi^+\left(\left.\frac{\partial}{\partial
\xi^\a}\right|_{\suup}\right)= \d
\Phi^-\left(\left.\frac{\partial}{\partial
\xi^\a}\right|_{\suup}\right), \label{eq:igualvectan} \] as can
the hypersurfaces $\sup^+\equiv \sup^-$, so henceforth we
represent both  $\sup^\pm$ by $\sup$.

We now define a 1-form $\bm n$, normal to the hypersurface $\sup$, as
\[
\bm n^\pm(\vec e^{\;\pm}_\a)=0.
\label{eq:normals}
\]
The vectors
$\{\vec n^{\;\pm},\vec e^{\;\pm}_\a\}$ constitute a basis
on the tangent spaces to $\mmm^\pm$ at $\sup^\pm$.
Since the first junction conditions allows the identification of
$\{\vec e^{\;+}_\a\}$ with $\{\vec e^{\;-}_\a\}$, we
only have to ensure that both bases have the same orientation and
that
$n^{+}_{i} n^{+i}\eqq n^{-}_{i} n^{-i}$ is satisfied
in order to identify the whole 4-dimensional
tangent spaces of $\mmm^\pm$ at $\sup$,
$\{\vec n^{\;+},\vec e^{\;+}_\a\}\equiv
\{\vec n^{\;-},\vec e^{\;-}_\a\}$.

The second fundamental forms are given by
\[
\label{secon}
H^\pm_{\a\b}=-n^\pm_i e^{\pm j}_\a\nabla^\pm_j e^{\pm i}_\b
\]
and the {\em second matching
conditions}, for non-null surfaces, are the equality of the second fundamental forms
\be
\label{secs}
H^+_{\a\b}=H^-_{\a\b}.
\ee
We note that these matching conditions do not depend on the choice of
the normal vectors.

\section{Cylindrically symmetric exterior and spatially homogeneous
interior}
We are interested in the particular cases for which the matching
surface inherits a certain symmetry of the two space-times
$(\mmm^\pm,g^\pm)$. Such matching is said to {\em preserve the
symmetry}. In practice one demands that the matching hypersurface
is tangent to the orbits of the symmetry group to be preserved
(see e.g. \cite{Vera01}). In the case we shall consider below, the
symmetry group is an Abelian $G_2$ and we are interested in the
matching of a cylindrical symmetric exterior to a spatially
homogeneous (and anisotropic) interior preserving the $G_2$
symmetry.

Since we want to preserve the symmetry of a $G_2$ exterior
space-time we must ensure that the interior also has the same
group of symmetries, possibly as a subgroup of a larger isometry
group. We wish to preserve an axial symmetry so that we require
the interior to have an axial symmetry in the sense of a periodic
Killing vector, vanishing on a symmetry axis. In the exterior
$\mmm^+$ the Killing vectors commute, so we require this in the
interior. Thus we require the interior to be spatially
homogeneous, with an axial symmetry and at least one Killing
vector commuting with the axial symmetry. This is sufficient to
force the interior to be an LRS space-time (see Mena, Tavakol \&
Vera \cite{MTV}).

We deal separately the metric forms for the interior and exterior.

\subsection{Interior}
All the LRS spatially homogeneous metrics can be written in the compact
form
\be
\label{compact}
ds^{2-}=-dt^2+a(t)^2 \bm{\theta}^2+b(t)^2\left[(dr-\epsilon rdz)^2+\Sigma(r)^2 d\varphi^2\right]
\ee
where
\[
\label{theta1}
\bm{\theta}=dz+n(F(r)+k)d\varphi
\]
and the functions $\Sigma$ and $F$ are given by
\[
\label{sigma}
\Sigma(r)=
\left\{
\begin{array}{cl}
\sin r,& k=+1\\
r, & k=0\\
\sinh r, & k=-1
\end{array}
\right.
~~
~~
and
~~
F(r)=
\left\{
\begin{array}{cl}
-\cos r,& k=+1\\
r^2/2, & k=0\\
\cosh r, & k=-1,
\end{array}
\right.
\]
and where $\epsilon$ and $n$ are given such that
\begin{equation}
\label{epsi}
\epsilon=0,1;~~n=0,1;~~\epsilon n=\epsilon k=0.
\end{equation}
We note that
\[
\label{Fprime}
\Sigma=F_{r};~~(\Sigma_{r})^2+k\Sigma^2=1,
\]
where a subscript denotes the
partial derivative with respect to the indicated variable.
The axial Killing vector is then given by
\[
\label{aija10}
\vec\eta_1=\partial_{\varphi},
\]
while the other three Killing vectors $\vec\eta_i, ~i=2$ to $4$ are
\bea
\label{aija3}
\vec\eta_2 &=&\p_z,\nonumber\\
\vec\eta_3 &=&\sin\varphi e^{\epsilon z}\p_r+
\cos\varphi(e^{\epsilon z}f(r)\p_\varphi+g(r)\p_z),\nonumber\\
\vec\eta_4 &=&\cos{\varphi}e^{\epsilon z}\p_r-\sin\varphi(e^{\epsilon z}
f(r)\p_\varphi+g(r)\p_z),\nonumber
\eea
where we have defined
$f(r)=\Sigma_{r}/\Sigma$ and $g(r)=n(\Sigma-f(F+k))$. The Killing vector which commutes with the axial Killing vector $\vec\eta_1$ is $\vec\eta_2$.
\begin{table}[!tb]
\begin{center}
\begin{tabular} {|c|c| c| c|}
\hline
KS/Bianchi types & $\epsilon$&$n$&$k$ \\
\hline
I &0&0&0
\\
KS &0&0&1
\\
KS,III & 0&0&-1
\\
IX & 0&1&1
\\
II & 0&1&0
\\
VIII,III &0&1&-1
\\
V,VII$_h$ &1&0&0
\\
\hline
\end{tabular}
\caption{\label{table:bianchis} Classification of the possible
$G_3$ on $S_3$ subgroups according to the values
of $\{\epsilon,k,n\}$ for the metric given by (\ref{compact}). KS denotes Kantowski-Sachs.}
\end{center}
\end{table}
We now assume a
perfect fluid interior and recall a result which is a direct consequence of the so-called
Israel conditions $n^{-i} T^-_{ij}\eqq n^{+i}T^+_{ij}$:
\begin{lemma}
Let $(\mmm,g)$ be a space-time resulting from the matching of two
space-times $(\mmm^\pm,g^\pm)$.
If $(\mmm^-,g^-)$ has a perfect fluid and $(\mmm^+,g^+)$ vacuum then
the perfect fluid has $p\eqq 0$. Furthermore, if $(\mmm^-,g^-)$ is spatially
homogeneous then $p=0$ everywhere.
\end{lemma}
The metric for the interior will be assumed to be given by
(\ref{compact}) with dust as matter content. For convenience we shall use
the form
\be
\label{non-static}
ds^{2-}=-A^2dt^2+B^2dr^2-2\epsilon r B^2 dr dz
+C^2 d{\varphi}^2
+2Edzd\varphi+D^2d{z}^2,
\ee
where $A,B,C,D$ and $E$ functions
of $t$ and $r$. The line-element (\ref{compact}) is recovered
by making the identifications
\bea
\label{idents}
A^2(t,r)&=&1,\nonumber\\
B^2(t,r)&=&b^2(t),\nonumber\\
C^2(t,r)&=&b^2(t)\Sigma^2(r,k)+n a^2(t)(F(r,k)+k)^2,\nonumber\\
D^2(t,r)&=&a^2(t)+\epsilon r^2 b^2(t),\nonumber\\
E(t,r)&=&n a^2(t) (F(r,k)+k).\nonumber
\eea
In the following we shall take the functions in (\ref{non-static})
to be arbitrary functions, with
$$
\epsilon E=0,
$$
which follows from (\ref{epsi}).
\subsection{Exterior}
The exterior will be assumed to be cylindrically symmetric vacuum
(with an Abelian $G_2$ on $S_2$). The most general metric form can
be written as \cite{BCM} \be \label{ext}
ds^{2+}=e^{2(\gamma-\psi)}(-dT^2+d\rho^2)+R^2e^{-2\psi}d\tilde\varphi^2+e^{2\psi}(d\tilde
z+Wd\tilde \varphi)^2 \ee where $\psi,\gamma,R,W$ are functions of
the coordinates $\rho,T$. For $W\ne 0$ the two Killing vectors are
not hypersurface orthogonal and the cylindrical gravitational
waves have two polarisations states. One of the EFEs is
(\ref{EFE2}) \[ R_{TT}-R_{\rho\rho}=0, \] which has the general
solution \be \label{Rext} R(T,\rho)=F(x)+G(y), \ee where
\[x=T+\rho;\;\;y=T-\rho.\]
The metric (\ref{ext}) is invariant under a
coordinate transformation \be \tau:(x,y)\to(f(x),g(y)),
\label{tau} \ee where $f$ and $g$ are arbitrary differentiable
functions (with non-zero derivative). One may use this freedom to
restrict $R$, but we shall make a different choice below, which is
to use $\tau$ to prescribe the matching surface $\sigma^+$. The
choices $W=0$ and $R=\rho$ in (\ref{ext}) lead to the well-known
Einstein-Rosen metric.

It will be important below to see when the $2$-surfaces of
transitivity are trapped or marginally-trapped. From (\ref{ext})
we see that these $2$-surfaces have intrinsic metric
$Rd\tilde\varphi d\tilde z$. Taking the two null-normals to these
$2$-surfaces, we see that such a $2$-surface is trapped if \be
4R_xR_y=(R_T-R_{\rho})(R_T+R_{\rho})\geq 0 \label{TS}, \ee and is
marginally-trapped if this expression is exactly zero. It follows
from (\ref{Rext}) that if $R_x=0$, at say $x=x_0$, then all
$2$-surfaces of transitivity with $x=x_0$ are marginally-trapped.
Similar statements follow with $y$ replacing $x$.

\subsection{The matching hypersurfaces}
The embeddings $\sigma^\pm$ can be defined by using coordinates on $\sigma$
denoted by
$\xi^\alpha=\{\lambda,\phi,\zeta  \}$. We choose $\phi$ such that
\[
d\Phi^-\left(\frac{\partial}{\partial
\phi}\right)=\left.\frac{\partial}{\partial
\varphi}\right|_{\sigma^-}=\vec e^{~-}_2.
\]
Since we perform the matching preserving the $G_2$ symmetry there must a
vector field $\vec\gamma$ which together with $\partial/\partial
\varphi|_{\sigma^-}$ generates the $G_2$ on $S_2$. In particular $\vec
\gamma$ must satisfy
\[
d\Phi^-(\vec \gamma)=c_1\left.\frac{\partial}{\partial \varphi}\right
|_{\sigma^-}+c_2\left.\frac{\partial}{\partial z}\right |_{\sigma^-}
\]
with $c_1,c_2$ constants such that $c_2\ne 0$ and otherwise arbitrary. We now use the
fact that $\partial/\partial \phi$ and $\vec \gamma$ must commute to choose $\vec\gamma=\partial/\partial \zeta$ and a coordinate
transformation $\zeta'=c_2\zeta,\phi'=\phi+c_1\zeta$ to get (after dropping
the primes):
\[
d\Phi^-\left(\frac{\partial}{\partial \zeta}\right)=\left.\frac{\partial }{\partial
z}\right |_{\sigma^-}=\vec e^{~-}_3.
\]
Finally, $\lambda$ is chosen such that $d\Phi^-(\partial/\partial
\lambda)$ is orthogonal to $\vec e^{~-}_2$ and $\vec e^{~-}_3$. This implies
\[
d\Phi^-\left(\frac{\partial}{\partial
\lambda}\right)=\frac{\partial \Phi^{0-}}{\partial
\lambda}\left.\frac{\partial}{\partial
t}\right|_{\sigma^-}+\frac{\partial \Phi^{1-}}{\partial
\lambda}\left.\frac{\partial}{\partial
r}\right|_{\sigma^-}+\left.\epsilon\Phi^{1-}\frac{B^2}{
D^2}\frac{\partial \Phi^{1-}}{\partial
\lambda}\frac{\partial}{\partial z}\right|_{\sigma^-}=\vec e^{~-}_1
\]
By denoting the embedding
$\{\Phi^{0-},\Phi^{1-},\Phi^{2-},\Phi^{3-}\}$
in (\ref{embed}) as $\{t,r,\varphi,z\}$,
the matching surface $\sigma^-$ is parametrized as
\[
\label{surf-}
\sigma^-=\{(t,r,\varphi,z): t=t(\lambda),
r=r(\lambda),\varphi=\phi,z=\zeta+f_{z}(\lambda) \},
\]
where $t(\lambda)$ and $r(\lambda)$
are functions of $\lambda$ restricted
by the fact that $d\Phi^-$ has to be of rank 3, that is
$\dot t^2+\dot r^2\neq 0$,
and
\be
\dot f_{z}(\lambda)\eqq\epsilon r(\lambda)\dot r(\lambda)\frac{B^2}
{D^2}
\label{dotfz}
\ee
where the dot denotes differentiation with respect to $\lambda$.

Consider now the embedding $\Phi^+$. The axial
killing vectors from both $\sigma^\pm$ must coincide at the matching
surface so
\[
d\Phi^+\left(\frac{\partial}{\partial
\phi}\right)=\left.\frac{\partial}{\partial
\tilde\varphi}\right|_{\sigma^+}=\vec e^+_2.
\]
and $d\Phi^+(\left.\partial/\partial \zeta)\right|_{\sigma^+}$ must complete the
basis of the $G_2$ on $S_2$. By a similar procedure as for $\sigma^-$ we can
always choose $\zeta$ and $\lambda$ such that
\bea
d\Phi^+\left(\frac{\partial}{\partial
\phi}\right)&=&\left.\frac{\partial}{\partial
\zeta}\right|_{\sigma^+}=\vec e^+_3\nonumber\\
d\Phi^+\left(\frac{\partial}{\partial
\lambda}\right)&=&\frac{\partial \Phi^{0+}}{\partial
\lambda}\left.\frac{\partial}{\partial
T}\right|_{\sigma^+}+\frac{\partial \Phi^{1+}}{\partial
\lambda}\left.\frac{\partial}{\partial
\rho}\right|_{\sigma^+}=\vec e^+_1 \nonumber\eea so that matching
surface $\sigma^+$ can be parametrized by \[
\sigma^+=\{(T,\rho,\tilde\varphi,\tilde
z):~T=T(\lambda),\rho=\rho(\lambda), \tilde \varphi=\phi,\tilde
z=\zeta\}.\nonumber \]
\section{The matching conditions}
In order to derive
the junction conditions we have to calculate the first and second
fundamental forms
for both $\sigma ^+$ and $\sigma^-$.
For the $g^-$ metric,
the parametric form of $\sigma^-$ gives
$dt|_{\sigma^-}=\dot{t}d\lambda,~ dr|_{\sigma^-}=\dot{r}d\lambda,~
d\varphi|_{\sigma^-}=d\phi$ and
$dz|_{\sigma^-}=d\zeta+\dot f_{z}d\lambda.$
Using (\ref{dotfz}), the first fundamental form
$\bar g^-_{ab}$
on $\sigma^-$ can be written as
\[
\label{hola1}
ds^{2-}|_{\sigma^-}\eqq (-A^2\dot{t}^2+{\cal B}^2\dot{r}^2)d\lambda^2
+{C}^2 d\phi^2+2{E}d\phi d\zeta
+{D}^2 d\zeta^2,
\]
where
$$
{\cal B}^2\equiv B^2\left(1-\epsilon r^2
\frac{B^2}{D}\right),
$$
Similarly the first fundamental form $\bar g^+$ on $\sigma^+$ is
given by \[ ds^{2+}|_{\sigma^+} \eqq e^{2(\gamma-\psi)}(-\dot
T^2+\dot
\rho^2)d\lambda^2+(R^2e^{-2\psi}+W^2e^{2\psi})d\phi^2+2We^{2\psi}d\zeta
d\phi+e^{2\psi}d\zeta^2 \] and then the first set of matching
conditions (\ref{aija5}) is: \bea \label{f1}
-A^2\dot t^2+ {\cal B}^2\dot r^2 &\eqq & e^{2(\gamma-\psi)}(-\dot T^2+\dot \rho^2)\\
C^2 & \eqq & R^2e^{-2\psi}+W^2e^{2\psi}\\
D^2 & \eqq & e^{2\psi}\\ \label{f4} E & \eqq & We^{2\psi} \eea The
normal forms to the matching surfaces can be taken as \bea
{\bm n}^-&=&A{\cal B}(-\dot rdt+\dot tdr)|_{\sigma^-}\nonumber\\
{\bm n}^+&=& e^{2(\gamma-\psi)}(-\dot \rho dT+\dot T
d\rho)|_{\sigma^+}\nonumber \eea
In our case, the interior space-time contains dust and so the
matching is  performed across a time-like hypersurface ruled by
matter trajectories, which are geodesics (this follows from the
matching conditions). Therefore, $\dot t =1$ and $\dot r =0$. From
the point of view of the exterior, the matching surface is given
parametrically as $(x(\lambda),y(\lambda))$ in terms of the
coordinates $x,y$ in (\ref{tau}). The product $\dot{x}\dot{y}$
does not vanish since the matching surface is everywhere time-like
and so we may exploit the transformation $\tau$ of (\ref{tau}) to
set $x=\lambda, y=\lambda$, or equivalently $\dot T=1$ and
$\dot\rho=0$. Now the matching surface is at $r=r_0, \rho=\rho_0$
and the time-coordinates can be taken to agree, $t=T$.

Under these circumstances we get the first set of matching
conditions (\ref{f1})-(\ref{f4}) in the form: \bea \label{gp}
\gamma &\eqq &\psi\\
b^2\Sigma^2+na^2(F+k)^2 &\eqq &R^2e^{-2\psi}+W^2e^{2\psi}\\
a^2+\epsilon r^2b^2 & \eqq & e^{2\psi} \\ \label{r4} na^2(F+k) &
\eqq & We^{2\psi}, \eea and the second set of matching conditions
(\ref{secs}) as \bea \label{m1}
\psi_\rho &\eqq& 0\\
\label{m2} \gamma_\rho &\eqq& 0\\ \label{m5}
\frac{na^2F_r}{b}&\eqq& e^{2\psi}W_\rho\\ \label{m4} \frac{C
C_r}{b}&\eqq& R R_\rho e^{-2\psi}+W W_\rho e^{2\psi}, \eea and for
$(DB)_t\ne 0$ \[ \epsilon \eqq 0. \] As a consequence of
(\ref{gp}) and $\dot T=1$ we also have \[ \label{m3} \gamma_T\eqq
\psi_T. \] The fact that $\epsilon=0$ excludes some Bianchi types:
\begin{prop}
A $G_4$ on $S_3$ LRS spacetime admitting a simply transitive
subgroup $G_3$ of Bianchi types $V$ or $VII_h$ cannot be matched
to a cylindrically symmetric non-stationary vacuum spacetime
across timelike surfaces preserving the cylindrical symmetry.
\end{prop}
We shall now look more closely at the cases with diagonal metrics.
\section{Diagonal cases}
For simplicity, we begin with a diagonal interior  metric, so that
$\epsilon=n=0$. Now $W$ and $W_{\rho}$ vanish on $\sigma$, from
(\ref{r4}) and (\ref{m5}). In each case below we first establish
that  $R\neq 0$ in the exterior, and we can then deduce that $W=0$
in the exterior, from uniqueness for the linear equation (\ref{W}) which it
satisfies. Assuming this for now,
the matching conditions give \bea \label{50}
e^\psi & \eqq & a\\
\label{mR}
R & \eqq & ab\Sigma \\
\label{mRr} R_\rho & \eqq & a\Sigma_r, \eea together with
$\gamma\eqq\psi$ and $\gamma_\rho\eqq\psi_\rho=0$. We first
consider the two explicit solutions corresponding to the flat
$(k=0)$ FLRW and to the Bianchi I cases and then consider implicit
solutions for Kantowski-Sachs and Bianchi III models. We may
summarise the results that we shall find in the following theorem.
\begin{theorem}
A cylindrical interior dust metric constructed from a collapsing
FLRW, Bianchi $I$ or Kantowski-Sachs cosmological model can be
matched to a diagonal vacuum exterior within the domain of
dependence ${\cal{D}}$ of the matching surface. The metric is
smooth up to the Cauchy horizon ${\cal{H}}$, but there are trapped
surfaces in the exterior. If the interior is a Bianchi $III$ model
(with space sections $\RR\times S^2$) then the matching can be
performed as far as the Cauchy horizons of the past and future
singularities provided the matter occupies less than half of the
$S^2$. If the matter occupies more than half of the $S^2$ then the
matching is at least possible near to the matching surface.
\end{theorem}
In Section 7 we shall present evidence that the exterior is in
fact singular on the Cauchy surface, and we shall suggest a
physical origin for this singularity.
\subsection{FLRW $k=0$}
We want the interior metric to be collapsing so that the relevant
solution to EFEs (\ref{diagonal}) is explicitly given by \be
\label{53} a(t)=b(t)=(\alpha-t)^{2/3}, \ee for $t\in
(-\infty,\alpha)$. The matching  surface $\sigma$ is the cylinder
$\rho=\rho_0,\, T<\alpha$, terminating in the singularity at
$T=\alpha$. At $\sigma$, by (\ref{mR}) and (\ref{mRr}) we have
\bea
R & \eqq & r_0(\alpha-T)^{4/3}\label{SR}\\
R_{\rho}&\eqq & (\alpha-T)^{\frac{2}{3}},\label{SRR} \eea
so that
in particular $R$ and $R_{\rho}$ on $\sigma$ are  positive for $T<\alpha$,
vanishing only at $T=\alpha$.

We shall determine the exterior using the matching conditions, in
the domain of  dependence ${\cal{D}}=\{T+\rho<\alpha+\rho_0,
\rho\geq\rho_0\}$ of $\sigma$ in ${\cal{Q}}$, when $\sigma$ is
taken to be at $\rho=\rho_0$. The boundary of the domain of
dependence, at $T+\rho=\alpha+\rho_0$, is the Cauchy horizon which
we shall call ${\cal{H}}$. It follows from the D'Alembert solution
of the one-dimensional wave equation for $R$ that, since the data
for it is positive, $R$ is positive in ${\cal{D}}$. (Now we may
conclude that $W=0$). We may obtain $R$ explicitly from
(\ref{SR}), (\ref{SRR}) and (\ref{Rext}) as \be
R=\frac{r_0}{2}(\alpha+\rho_0-T-\rho)^{\frac{4}{3}}-\frac{3}{10}(\alpha+\rho_0-T-\rho)^{\frac{5}{3}}+\frac{r_0}{2}(\alpha+\rho_0-T+\rho)^{\frac{4}{3}}+
\frac{3}{10}(\alpha+\rho_0-T+\rho)^{\frac{5}{3}}. \ee From this
formula for $R$ we can investigate  the condition (\ref{TS}) for
trapped-ness. On $\sigma$ it is easy to see that $R_y$ vanishes
only at $T=\alpha$, while $R_x$ vanishes at $T=\alpha$ and at
$T:=T_0=\alpha-\left(\frac{8r_0}{3}\right)^{3/2}$. Thus there is a
marginally-trapped cylinder at $T=T_0$ on $\sigma$ and this gives
rise in the exterior to whole null hypersurface ${\cal{N}}_0$ of
marginally-trapped cylinders at $T+\rho=T_0+\rho_0$. Cylinders to
the past of ${\cal{N}}_0$ are not trapped and those to the future
are.

We have seen that $R$ is non-negative in ${\cal{D}}$  and by
inspection it vanishes only in the corner at $T=\alpha,
\rho=\rho_0$, and is smooth in the interior of ${\cal{D}}$, though
not along the Cauchy horizon ${\cal{H}}$ at $T+\rho=\alpha+\rho_0$
where second and higher derivatives diverge.

At $\sigma$, by (\ref{m1}), (\ref{50}) and (\ref{53}) we have \bea
\psi & \eqq & \frac{2}{3}\ln{(\alpha-T)}\nonumber\\
\psi_{\rho} & \eqq & 0. \nonumber\eea This is the data for $\psi$
in the exterior, and  it is smooth (in fact analytic) up to
$T=\alpha$. Given what we know about $R$, it now follows from
standard theorems for linear hyperbolic equations (see e.g.
section $6.5$ in \cite{tay}) that a unique $\psi$ exists satisying
(\ref{wave}) in the interior of ${\cal{D}}$, with the given data
on $\sigma$. There is no reason to expect the solution to be
well-defined at ${\cal{H}}$, since it diverges on $\sigma$ at
$T=\alpha$, and in fact we shall argue in section 7 that there is
a curvature singularity along ${\cal{H}}$.

Since $\gamma\eqq\psi$ and $\gamma_\rho\eqq\psi_\rho$, once  we
have $\psi$ in ${\cal{D}}$ the same argument applied to
(\ref{gammapsi}) gives existence and uniqueness for $\gamma$ in
${\cal{D}}$, and we have constructed a vacuum exterior which
matches on to the dust interior, at least inside ${\cal{D}}$. (In this and all the examples, the constraint equations (\ref{gammar}) and (\ref{gammaT}) are identically satisfied by virtue of the field equations in the interior, as must be the case.)
\subsection{Bianchi I}
The collapsing solution to the EFEs (\ref{diagonal}) in the
interior can be taken to be \bea
a(t)&=&(\alpha-t)^{-1/3}(\beta-t)\nonumber\\
b(t)&=&(\alpha-t)^{2/3} \nonumber\eea for $t<\rm{
min}\{\alpha,\beta\}$ and $r\le r_0$. If $\alpha =\beta$ this
reduces to the previous case. The character of the solution
depends on which is the smaller of $\alpha$ and $\beta$. If
$\alpha<\beta$ then $a$ diverges and $b$ vanishes at the
singularity. This is the behaviour of the Kasner (vacuum) solution
and we can call it Kasner-like. If $\alpha=\beta$ then $a$ and $b$
vanish at the same rate, which is precisely the FLRW behaviour so
that we may call it FLRW-like. If $\alpha>\beta$ then $a$ vanishes
but $b$ does not and we call this disk-like.

Since the interior is given explicitly we can use the matching
conditions as before to get the exterior functions. We again  find
that the data $R$ and $R_{\rho}$ are positive so that $R$ is
positive in ${\cal{D}}$ and $W=0$. Explicitly, (\ref{mR}),
(\ref{mRr}), (\ref{Rext}) imply \bea
R&=&\frac{r_0}{2}(\alpha+\rho_0-T-\rho)^{\frac{1}{3}}(\beta+\rho_0-T-\rho)-\frac{3}{4}(\alpha+\rho_0-T-\rho)^{\frac{2}{3}}(\beta+\rho_0-T-\rho)\nonumber\\
&&+
\frac{9}{20}(\alpha+\rho_0-T-\rho)^{\frac{5}{3}}
\nonumber\\
&&+\frac{r_0}{2}(\alpha-\rho_0-T+\rho)^{\frac{1}{3}}(\beta-\rho_0-T+\rho)+\frac{3}{4}(\alpha-\rho_0-T+\rho)^{\frac{2}{3}}(\beta-\rho_0-T+\rho)\nonumber\\&&-
\frac{9}{20}(\alpha+\rho_0-T+\rho)^{\frac{5}{3}}. \nonumber\eea It
is straightforward  to see that $R_y$ is always negative on
$\sigma$ while $R_x$ necessarily changes sign.  Thus, by
(\ref{TS}), there is always a marginally-trapped cylinder on
$\sigma$ and this gives rise to a null hypersurface of
marginally-trapped cylinders ${\cal{N}}_0$ as in the previous case
(for suitable combinations of the parameters $\alpha, \beta$ and
$\rho_0$ there may be three such hypersurfaces). However $R$ is
positive in the exterior, vanishing only in the corner
$\rho=\rho_0,\, T=\rm{min}(\alpha, \beta)$. Note also that $R_x$
diverges at the singularity, and therefore on the Cauchy horizon,
if the singularity is Kasner-like.

For the other exterior metric functions we get from (\ref{50}) \[
\psi\eqq\gamma\eqq \ln {(\beta-T)}-\frac{1}{3}\ln{(\alpha-T)} \]
and with $\gamma_{\rho}\eqq\psi_{\rho}\eqq 0$ we  again have
existence and uniqueness of solutions in the interior of
${\cal{D}}$. As before, we expect the solution to diverge near
${\cal{H}}$, a point we shall return to below. We have then
constructed a vacuum exterior which matches on to the Bianchi I
dust interior, at least inside ${\cal{D}}$.
\subsection{Kantowski-Sachs and Bianchi III}
For these interiors, implicit solutions are  known and we give
them in the appendix. Dealing first with the Kantowski-Sachs
metric, the data for $R$ are positive on $\sigma$ so that as usual
$R$ is positive in the exterior, out to the Cauchy horizon, and
$W=0$. The singularity is Kasner-like, FLRW-like or disk-like
according as $C_0$ is positive, zero or negative in (\ref{KS-1}).
In each case, the quantity $R_T-R_{\rho}$ changes sign from
positive in the infinite past to negative near the singularity so
that there are trapped cylinders on $\sigma$ near the singularity
and these persist in the exterior as before. The existence
argument carries through just as before, out to the Cauchy
horizon.

There are two extra complications with the  Bianchi III metric.
The first is that there are singularities in the past as well as in the
future. The second is that $\Sigma = \sin{r}$ so that
$R_{\rho}\eqq a\cos{r_0}$ which can be negative or zero.

Suppose the past and future singularities lie  on $\sigma$ at
$T=T_0$ and $T=T_1$ respectively, then the domain of dependence is
restricted to
\[{\cal{D}}=\{T+\rho<T_1+\rho_0,\,T-\rho>T_0-\rho_0\}\]
and there is a past Cauchy horizon as well as a  future one. The
nature of the singularities depends on the relative sizes of the
constants $M$ and $C_0$ in (\ref{KS1}) and various combinations of
Kasner-like, disk-like and FLRW-like are possible. However, at
least one of the past and future singularities is always
Kasner-like.

Now for $R$, suppose first that $0<r_0\leq\pi/2$,  then the data
for $R$ are positive so $R$ is positive in the exterior and $W=0$.
The proof of existence is as before. On the other hand, if
$\pi/2<r_0<\pi$ then $R_{\rho}$ is negative on $\sigma$ and so $R$
may vanish inside ${\cal{D}}$. If $R$ vanished in a smooth way,
this could just be the sign of a (second) regular axis for the
axial Killing vector. To see that this is not the case we recall
from the discussion of the Bianchi I case above that the
derivative of $R$ is singular on the Cauchy horizon emanating from
a Kasner-like singularity. Since one of the singularities must be
Kasner-like, $R$ cannot be smooth at any axis meeting that part of
the Cauchy horizon.

For each kind of singularity there are both
future-marginally-trapped and past-marginally-trapped cylinders on
$\sigma$. This follows from an analysis of the implicit solution
(\ref{KS1}), but can be seen just by considering asymptotic forms
near the different kinds of singularity as they have already
occurred. The global picture is complicated by the presence of
both past and future Cauchy horizons, but the existence of the
exterior in the whole domain of dependence follows as before if
$r_0\leq\pi/2$ and follows near to $\sigma$ if $r_0>\pi/2$.
\section{A Non-diagonal Example}
This case is much harder and we have only  partial results. In
principle, one could repeat the work of the previous section with
the non-diagonal interior metrics of Bianchi types II, VIII and IX
matched to non-diagonal exterior vacuum metrics. This is hampered
by two things: the non-existence, in the literature at least, of
even implicit solutions for the interior metric; and the fact that
the EFEs are now coupled and cannot be solved one by one.

For the interior solution, the Bianchi type II  has been taken
furthest and we shall consider just some particular instances of
this case. The constants in (\ref{compact}) are $\epsilon=k=0$ and
$n=1$ so that the matching conditions from (\ref{gp})-(\ref{m4})
are
\bea
\psi&\eqq &\log a\nonumber\\
R&\eqq&abr_0\nonumber\\
W&\eqq&r_0^{\,2}/2\nonumber\\
\psi_{\rho}&\eqq&0\nonumber\\
R_{\rho}&\eqq&a\nonumber\\
W_{\rho}&\eqq&r_0/b \nonumber \eea
together, as always, with $\gamma\eqq\psi$ and $\gamma_{\rho}\eqq\psi_{\rho}$.

The field equations for the interior metric are given by
(\ref{BII}). They have been solved in terms of $\mu$ subject to a
single ODE by Maartens and Nel \cite{MN} as follows:
\bea
a(t)&=&\mu\exp(-2\int^t\mu(s)(\alpha-s)d s)\nonumber\\
b(t)&=&\mu^{-1}\exp(\int^t\mu(s)(\alpha-s)d s) \nonumber\eea
where $\mu(t)$ satisfies
\be
\mu_{tt}-\frac{4}{\mu}(\mu_t)^2+8(\alpha-t)\mu\mu_t-\frac{1}{2}\mu^2-6(\alpha-t)^2\mu^3=0.
\label{mu}
\ee
In (\ref{mu}), $\alpha$ is an additive  constant and the
singularity will be at $t=\alpha$. It is possible to reduce the
order of the equation for $\mu$, but it does not seem possible to
find a general solution. Two particular solutions are given by
$\mu(t)=k/(\alpha-t)^2$ with $k=5/4$ or $k=4/3$, and now $a$ and
$b$ are just powers of $(\alpha-t)$.

Proceeding as before, we find that $R$ is strictly  positive, and
in fact analytic, in the domain of dependence. There are trapped
surfaces on $\sigma$ between the singularity at $T=\alpha$ and a
value $T_0$ which depends on the constants of integration. Thus
there are trapped surfaces in the exterior.

To obtain the other functions appearing in the exterior metric  we
need to solve the system of equations
(\ref{gammapsi})-(\ref{gammaT}). This is a much-studied system and
a good deal is known. Local existence in a closely related setting
was shown by Chrusciel \cite{CH}. To obtain a more global result,
and in particular existence out to the Cauchy horizon, the methods
of Berger et al. \cite{BCM} would be expected to work, but we have
not done this.


\section{Properties of the solutions}


We shall concentrate on the diagonal cases,  where existence is
known. We have found that a radiating exterior can be matched to a
collapsing interior but there are always trapped surfaces in the
exterior. From the work of Berger et al. \cite{BCM} this means
that the solutions cannot be asymptotically flat, so that we have
the following corollary to Theorem 1:
\begin{corr}
Asymptotically flat vacuum cylindrically symmetric spacetimes
cannot be matched to the boundary of solid cylinders in LRS
spatially homogeneous perfect fluid spacetimes with diagonal
metrics, preserving the cylindrical symmetry.
\end{corr}
We now want to present some evidence that these solutions are
actually singular on the Cauchy horizon ${\cal H}$. We begin by
introducing the quantity $u$ defined by
\[u=R^{1/2}\psi,\]
then (\ref{wave}) becomes
\[u_{xy}=\frac{R_xR_y}{4R^2}u.\]
It follows from this equation that  if $R_x=0, R\neq 0$ at some
value $x=x_0$ then $u_x$ is constant on $x=x_0$, and therefore so
is $R^{1/2}\psi_x$. In the FLRW case, precisely this happens on
${\cal H}$ so that $R^{1/2}\psi_x$ is constant on  ${\cal H}$.
However this quantity diverges in the corner where  ${\cal H}$
meets the matching surface, and it is therefore singular on
${\cal H}$. Once $\psi$ is not differentiable at  ${\cal H}$,
there is no reason to expect $\gamma$ to even be finite there, in
which case the metric is singular.

One can alternatively look at curvature  components, specifically
at the Weyl spinor in the exterior. We introduce the null tetrad
\bea
l&=&2^{-1/2}{\rm e}^{\psi-\gamma}(\partial_T+\partial_\rho)\nonumber\\
n&=&2^{-1/2}{\rm e}^{\psi-\gamma}(\partial_T-\partial_\rho)\nonumber\\
m&=&2^{-1/2}({\rm
e}^{-\psi}\partial_z+\rm{i}\rm{e}^{\psi}\partial_\phi),\nonumber
\eea
and calculate the (nonzero) components of the Weyl spinor in this tetrad as
\bea
\Psi_0&=&-2{\rm e}^{2(\psi-\gamma)}(\psi_{xx}+\frac{1}{R}R_x\psi_x-\frac{1}{2R}R_{xx})\nonumber\\
\Psi_2&=&2{\rm e}^{2(\psi-\gamma)}(\psi_{xy}-\gamma_{xy})\nonumber\\
\Psi_4&=&-2{\rm
e}^{2(\psi-\gamma)}(\psi_{yy}+\frac{1}{R}R_y\psi_y-\frac{1}{2R}R_{yy}).\nonumber
\eea
If we restrict these quantities to $\sigma$ for the case of an
FLRW interior, then we find that $\Psi_0=\Psi_4=0$ but that
$\Psi_2$ diverges as the singularity is approached. This is in the
corner where  ${\cal H}$ meets $\sigma$ and then we expect this
singularity to propagate along the Cauchy horizon.

We also note the appearance of the term $R_{xx}$ in $\Psi_0$,
which blows up  at  ${\cal H}$ for all our examples. We cannot
explicitly evaluate the other terms in $\Psi_0$, but this suggests
that $\Psi_0$ diverges at  ${\cal H}$. This component of the Weyl
spinor, in this tetrad, is interpreted as incoming radiation,
which therefore becomes singular at the Cauchy horizon. This
interpretation leads us to a physical explanation of the exterior
singularity: suppose at a particular instant we set up Cauchy data
with a spatially homogeneous interior, and we want the interior to
remain spatially homogeneous even as it collapses. This requires
incoming radiation and these examples show that it requires more
and more as the interior approaches the singularity, leading to a
singularity in the exterior on the Cauchy horizon.
\begin{figure}[!htb]
\centerline{\def\epsfsize#1#2{0.8#1}\epsffile{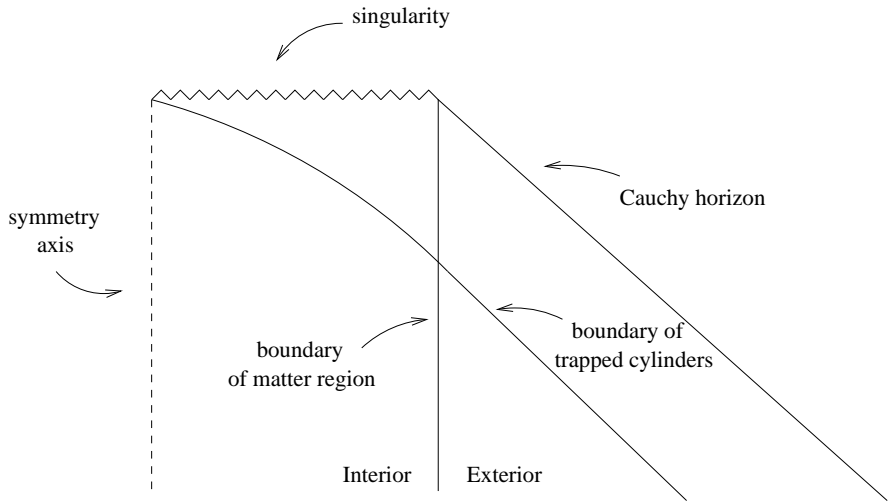}}
\caption{\label{fig} Schematic diagram of the spacetime
structure.}
\end{figure}
\vspace{.2in}

\centerline{\bf Acknowledgments} \vspace{.2in} FCM thanks Ra\"ul
Vera and Brien Nolan for usefull discussions, Funda\c{c}\~ao
Calouste Gulbenkian for grant 21-58348-B and Centro de
Matem\'atica, Universidade do Minho, for support.
\section*{Appendix}
The EFEs for the cylindrically symmetric vacuum metric (\ref{ext}) are
\bea \label{gammapsi}
0&=&\gamma_{TT}-\gamma_{\rho\rho}-\psi_\rho^2+\psi_T^2-\frac{e^{4\gamma}}{4R^2}((W_T)^2-(W_{\rho})^2)\\
\label{EFE2}
0&=&R_{TT}-R_{\rho\rho}\\
\label{wave}
0&=&\psi_{TT}+\frac{R_T}{R}\psi_T-\psi_{\rho\rho}-\frac{R_\rho}{R}\psi_\rho-\frac{e^{4\gamma}}{2R^2}((W_T)^2-(W_{\rho})^2)\\
\label{W}
0&=&W_{TT}-W_{\rho\rho}-\frac{R_T}{R}W_T+\frac{R_\rho}{R}W_{\rho}+4W_T\gamma_T-4W_{\rho}\gamma_{\rho}
\eea together with the two constraint equations \bea
\label{gammar} \gamma_\rho & = & \frac{1}{R_\rho^2-
R_T^2}(RR_\rho(\psi_T^2+\psi_\rho^2)-2RR_T\psi_T\psi_\rho+R_\rho
R_{\rho\rho}-R_TR_{T\rho})\\
\label{gammaT} \gamma_T & = & \frac{1}{R_T^2-
R_\rho^2}(RR_t(\psi_T^2+\psi_\rho^2)-2RR_\rho\psi_T\psi_\rho+R_T
R_{\rho\rho}-R_\rho R_{T\rho}). \eea The EFEs for LRS spatially
homogeneous diagonal dust metrics (\ref{compact}) are \bea
\mu &=& 2\frac{a_t b_t}{ab}+\frac{b_t^2}{b^2}+\frac{k}{b^2}\nonumber\\
\label{diagonal}
0 &=& 2\frac{b_{tt}}{b}+\frac{b_t^2}{b^2}+\frac{k}{b^2}\\
0 &=& \frac{b_{tt}}{b}+\frac{a_t
b_t}{ab}+\frac{a_{tt}}{a}\nonumber \eea The solution for $k=1$
(Bianchi III) in parametric form is \cite{Exact} \bea \label{KS1}
a \cos \eta & =& M(\eta\sin\eta+\cos\eta)+C_0\sin\eta\\
b&=&C_1\cos^2\eta,~~dt=2b d\eta \nonumber \eea
where $M$ and $C_1$ are positive constants.\\
\\\\
The solution for $k=-1$ (Kantowski-Sachs) in parametric form is
\cite{Exact} \bea \label{KS-1}
a \sinh \eta & =&\left[M(\eta\cosh\eta-\sinh\eta)+C_0\cosh\eta\right]\\
b&=&C_1\cosh^2\eta,~~dt=2bd\eta \nonumber \eea with $M$ and $C_1$
positive constants.
\\\\
The EFEs for LRS spatially homogeneous non-diagonal dust metrics
(\ref{compact}) of Bianchi II are \bea
\mu&=&\frac{b_t^2}{b^2}+2\frac{a_tb_t}{ab}-\frac{a^2}{4b^4}\nonumber\\
0&=&2\frac{b_{tt}}{b}+\frac{b_t^2}{b^2}-\frac{3a^2}{4b^4}\label{BII}\\
0&=&\frac{b_{tt}}{b}+\frac{a_{tt}}{a}+\frac{a_tb_t}{ab}+\frac{a^2}{4b^4}\nonumber
\eea from which the solution of Maartens and Nel \cite{MN} given
in the text can be found.

\end{document}